\documentclass{article}

\usepackage{booktabs}
\usepackage{graphics}
\usepackage{graphicx}
\usepackage{listings}
\usepackage{nameref}
\usepackage{algorithm}
\usepackage{algpseudocode}
\algrenewcommand\algorithmicrequire{\textbf{Input:}}
\algrenewcommand\algorithmicensure{\textbf{Output:}}
\usepackage{bm}
\usepackage{amsmath}
\usepackage{amssymb}
\usepackage{caption} 
\usepackage{multirow}
\usepackage{color}
\usepackage[table]{xcolor}

\usepackage{arxiv}

\usepackage[utf8]{inputenc} 
\usepackage[T1]{fontenc}    
\usepackage{hyperref}       
\usepackage{url}            
\usepackage{booktabs}       
\usepackage{amsfonts}       
\usepackage{nicefrac}       
\usepackage{microtype}      
\usepackage{lipsum}
\usepackage{graphicx}
\graphicspath{ {./images/} }

\begin{document}
\title{Accurate and efficient protein embedding using multi-teacher distillation learning}

\author{
 Jiayu Shang \\
  Dept. of Electrical Engineering\\
  City University of Hong Kong\\
  Kowloon, Hong Kong SAR, China\\
  \And
 Cheng Peng \\
  Dept. of Electrical Engineering\\
  City University of Hong Kong\\
  Kowloon, Hong Kong SAR, China\\
  \And
 Yongxin Ji \\
  Dept. of Electrical Engineering\\
  City University of Hong Kong\\
  Kowloon, Hong Kong SAR, China\\
    \And
   Jiaojiao Guan \\
  Dept. of Electrical Engineering\\
  City University of Hong Kong\\
  Kowloon, Hong Kong SAR, China\\
    \And
   Dehan Cai \\
  Dept. of Electrical Engineering\\
  City University of Hong Kong\\
  Kowloon, Hong Kong SAR, China\\
  \And
 Xubo Tang \\
  Dept. of Electrical Engineering\\
  City University of Hong Kong\\
  Kowloon, Hong Kong SAR, China\\
  \And
 Yanni Sun \\
  Dept. of Electrical Engineering\\
  City University of Hong Kong\\
  Kowloon, Hong Kong SAR, China\\
}

\maketitle
\begin{abstract}
\textbf{Motivation:}  Protein embedding, which represents proteins as numerical vectors, is a crucial step in various learning-based protein annotation/classification problems, including gene ontology prediction, protein-protein interaction prediction, and protein structure prediction. However, existing protein embedding methods are often computationally expensive due to their large number of parameters, which can reach millions or even billions. The growing availability of large-scale protein datasets and the need for efficient analysis tools have created a pressing demand for efficient protein embedding methods. \\
\textbf{Results:} We propose a novel protein embedding approach based on multi-teacher distillation learning, which leverages the knowledge of multiple pre-trained protein embedding models to learn a compact and informative representation of proteins. Our method achieves comparable performance to state-of-the-art methods while significantly reducing computational costs and resource requirements. Specifically, our approach reduces computational time by \textasciitilde70\% and maintains $\pm$1.5\% accuracy as the original large models. This makes our method well-suited for large-scale protein analysis and enables the bioinformatics community to perform protein embedding tasks more efficiently.\\
\textbf{Availability:} The source code od MTDP is available via \href{https://github.com/KennthShang/MTDP}{https://github.com/KennthShang/MTDP}\\
\textbf{Contact:} \href{yannisun@cityu.edu.hk}{yannisun@cityu.edu.hk}\\
\end{abstract}

\section{Introduction}
\label{sec:intro}
Protein characterization and annotation provide the foundational knowledge necessary to unravel the complex mechanisms underlying many biological processes. 
However, the complexity and variability of protein sequences pose significant challenges to traditional analysis methods, which struggle to capture their intricate patterns and relationships. Recently, deep learning-based algorithms have demonstrated remarkable success in protein data analysis, leveraging their ability to learn complex patterns and relationships from large datasets to predict gene ontology, 2D-/3D-structure, and protein stability \cite{hwang2024genomic, thumuluri2022deeploc, fang2023deepprosite}. An essential requirement for these successful applications is the representation of protein sequences as fixed-length numerical vectors, which encapsulate sequence composition information, structural signatures, and evolutionary relationships. Protein embedding models, which aim to extract informative and effective representations of proteins, play a crucial role in this process.

Recent advances in protein embedding models, which leverage unsupervised learning strategies such as masked language modeling, have yielded promising results in various bioinformatics applications \cite{rives2021biological, elnaggar2021prottrans}. However, as researchers strive to improve performance, these models have become increasingly large and complex, with millions or even billions of parameters, leading to a significant increase in computational resources required, particularly high-performance GPUs, and often necessitating lengthy running times. For example, processing a \textasciitilde1Mb FAA file using a large model on a 24Gb memory GPU card can take approximately 7 hours, thereby hindering research progress. In addition, the rapid growth of bioinformatics data containing sheer amount of protein sequences further exacerbates this issues. Thus, there is a growing demand for efficient protein embedding methods that balance efficiency and performance.

\begin{figure*}[h!]
    \centering
    \includegraphics[width=0.8\linewidth]{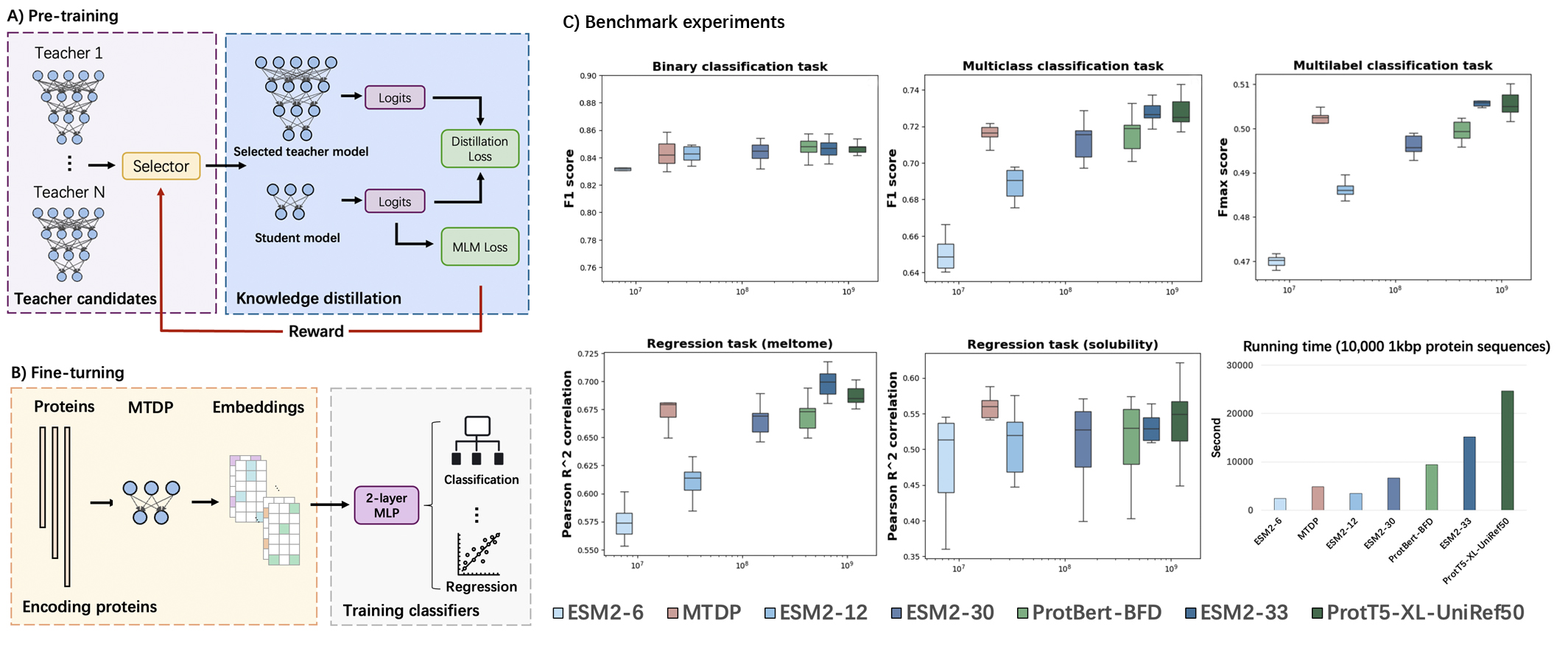}
    \caption{A) The framework of multi-teacher distillation; B) Extracting protein embeddings using MTDP and use the embeddings to train classifiers; C) The benchmark experiments on a wide range of protein-related tasks. Y-axis: the metric in each task. X-axis: the number of parameters in the model. }
    \label{fig:figure}
\end{figure*}

In this study, we propose MTDP, a \textbf{M}ulti-\textbf{T}eacher \textbf{D}istillation approach for \textbf{P}rotein embedding, which aims to enhance efficiency while preserving high-resolution representations. By leveraging the knowledge of multiple pre-trained protein embedding models, MTDP learns a compact and informative representation of proteins. It takes amino acid sequences as input and generates embeddings that encode biologically meaningful features of protein structure and function. We evaluated MTDP in various scenarios and benchmarked it against widely used large protein embedding models. Our results demonstrate that MTDP achieves state-of-the-art performance while significantly reducing computational costs and resource requirements.

\section{Methods and materials}
\label{sec:method}

Knowledge distillation is a training method used to transfer knowledge from a large, pre-trained model (the teacher) to a small, efficient model (the student). This approach can significantly reduce computational costs and memory requirements, as many parameters in deep neural networks can be redundant or even ineffective \cite{gromov2024unreasonable}. Recently, researchers have applied knowledge distillation to large protein language models to improve efficiency in protein embedding. However, traditional knowledge distillation methods often face two significant challenges. First, the student model's performance typically fails to match the teacher model's performance. For instance, DistilProtBert \cite{geffen2022distilprotbert}, a distilled protein language model, has shown promise in protein embedding but still struggles to match its teacher models' performance. Second, selecting a suitable teacher model can be a daunting task. As different teacher models may excel in different aspects of protein representation, the choice of teacher models can be a key factor that leads to inferior performance of the student model.

To address these challenges, we propose MTDP, a multi-teacher distillation approach that leverages the strengths of multiple pre-trained protein language models. During training, MTDP adaptively selects the most informative teachers for each training sample, allowing the student model to learn a more comprehensive representation of proteins. By aggregating and distilling knowledge from multiple teachers, this approach yields a more accurate and efficient protein embedding method. In the following section, we provide a detailed description of MTDP and outline the datasets used for pre-training and fine-tuning our model.

\subsection{Model structure}
The MTDP model leverages the ESM \cite{rives2021biological} and ProtTrans-series \cite{elnaggar2021prottrans} Transformers for distillation, which have demonstrated state-of-the-art performance in various protein-related analysis tasks. As a student model, MTDP is a six-layer T5 transformer model, well-suited for handling diverse embedding processing tasks within a unified framework. Following the trend that larger models tend to perform better, we select ESM2-33 and ProtT5-XL-UniRef50 as our teacher models, which are the largest models that can be run on a 24Gb commercial GPU, commonly found in workstations and medium-level high-performance computing systems.  Notably, the MTDP model has a significantly smaller parameter scale (\textasciitilde20 million) compared to ESM2-33 (\textasciitilde650 million) and ProtT5-XL-UniRef50 (\textasciitilde120 million), at approximately 3\% and 1.6\% of their sizes, respectively.

As illustrated in Fig. \ref{fig:figure} A, during pre-training, MTDP employs an adaptive teacher selection mechanism for knowledge distillation when pre-training. This is achieved through a reinforcement learning framework, which which learns a policy to select the optimal teacher model for each sample. The policy is trained using a reward function that evaluates the quality of the student model's output, with the objective of maximizing the expected cumulative reward over the training process. Because the scheduling policy used in our paper is a standard algorithm used in machine learning filed, we refer readers to \cite{yuan2021reinforced} for a detailed explanation of the reinforcement learning framework. 

Our MTDP model is trained using the masked language modeling task, where a random subset of amino acids in the protein sequence is masked, and the model predicts the original amino acids. The training objective consists of two components: the masked language model loss, which measures the difference between the predicted and original amino acids, and the distillation loss, which measures the difference between the student model's output and the teacher models' outputs. The distillation loss is calculated using the Kullback-Leibler (KL) divergence, which encourages the student model to mimic the teacher models' behavior. A detailed description of calculating losses can be found in the supplementary file. By jointly optimizing these two losses, our MTDP approach enables the student model to learn a compact and informative representation of proteins, effectively leveraging the strengths of multiple teacher models

\subsection{Data}
\subsubsection{Pre-training data}
MTDP is pre-trained on \textasciitilde500,000 proteins from UniProtKB (Swiss-Prot) provided by the teacher model \cite{elnaggar2021prottrans}. To prevent excessive memory usage, we followed the design of other protein embedding models and set 1,000 as the maximum length. Specifically, if the proteins exceed 1,000 amino acids in length, we only consider the first 1,000 amino acids. We employ an offline training strategy, utilizing ESM2-33 and ProtT5-XL-UniRef50 to generate protein embeddings on the same data with provided parameters. The model is pre-trained on the masked language task with a masking probability of 15\%. We conduct the pre-training process on four RTX 3080 24 GB Nvidia GPUs, with a batch size of 16. We use the AdamW optimizer with a learning rate of 3e-4. The model is trained for ten epochs using mixed precision with a warmup ratio of 0.1. To demonstrate the advantage of multi-teacher distillation, we also train two distillation models, each using only ESM2-33 or ProtT5-XL-UniRef50 as the teacher, and compare their performance with our MTDP model.

\subsubsection{Fine-tuning data}
We evaluate MTDP on several benchmark tasks, with dataset information summarized as follows. All these datasets are widely used for the evaluation of protein embedding model \cite{elnaggar2021prottrans, rives2021biological, outeiral2024codon} and detailed information of the datasets can be found in the supplementary.

\begin{itemize}
\item[$\bullet$] \textit{Binary classification}: classify proteins into membrane-bound or water-soluble.
\item[$\bullet$] \textit{Multiclass classification}: classify proteins into ten classes of subcellular localization.
\item[$\bullet$] \textit{Multi-label classification}: protein function prediction with gene ontology.
\item[$\bullet$] \textit{Regression task}: predict protein stability using (i) melting temperatures and (ii) solubility databases.
\end{itemize}

As shown in Fig. \ref{fig:figure} B, during fine-tuning, we follow the standard setting and apply global average pooling to the MTDP output to generate per-protein embeddings. Then, for each task, we feed these embeddings into a simple two-layer MLP and minimize the loss using the AdamW optimizer. We employ cross-entropy loss for classification tasks and mean squared error for regression tasks.

\section{Result}
\label{sec:result}
We benchmarked MTDP against six commonly used protein language models from the ESM-series (ESM2-6, ESM2-12, ESM2-30, and ESM2-33) and ProtTrans-series (ProtBert-BFD and ProtT5-XL-UniRef50). We selected these models because they can at least encode a single protein at a time on a 24 Gb GPU unit, making them suitable for comparison with our MTDP model. In the ProtTrans-series, there are also models with the same structure but trained on different datasets. In this case, we chose the best-performing model reported in their paper \cite{elnaggar2021prottrans} as the benchmark model.

In each task, we employed ten-fold cross-validation, recording performance metrics to generate box plots. Detailed descriptions of the metrics for each task are available in the supplementary file. Additionally, we documented the model size and running time for each model to aid in comparative analysis.

\subsection{Evaluation tasks}
As shown in Fig. \ref{fig:figure} C, by distilling knowledge from the teacher models, our MTDP approach learns a compact and informative representation of proteins, striking a balance between performance and efficiency. The results reveal that MTDP achieves nearly the same performance as its teachers, ESM2-33 and ProtT5-XL-UniRef50, across different tasks. Moreover, MTDP even outperforms its teachers in some tasks. Furthermore, the ablation study shown in Supplementary Fig. S1 demonstrates that MTDP's performance is significantly higher than that of a student model trained with a single teacher, highlighting the benefits of the multi-teacher distillation framework.

\subsection{Efficiency of MTDP}
Rapid analysis and interpretation of protein sequences are crucial, particularly when dealing with large-scale protein data. In our experiments, we ran all models on a single RTX 3080 24 GB Nvidia GPU and recorded the running time for processing 10,000 1kbp protein sequences. The results demonstrate that MTDP can encode protein sequences at very high speeds with almost no sacrifice in performance. Additionally, it is noteworthy that MTDP, ESM2-6, and ESM2-12 can be deployed on smaller GPU memories (e.g., RTX 2080Ti 11Gb), thereby expanding their range of usage scenarios.

\section{Discussion}

In this paper, we present MTDP, an accurate and efficient protein embedding method. We demonstrate the reliability of MTDP in various protein-related analysis scenarios. Notably, MTDP achieves high-resolution protein embeddings comparable to those of large protein language models while significantly reducing computational costs and resource requirements, making it a promising approach for large-scale protein analysis.

Although MTDP has significantly improved protein embedding, there are still some limitations. Notably, the current version of MTDP is only trained on UniProtKB (Swiss-Prot), which is a small subset of the UniRef50 dataset widely used for training large models like ESM2-33 and ProtT5-XL-UniRef50. To address this limitation, we plan to continuously update the UniRef50 and UniRef100 versions of MTDP in our GitHub repository. We will also provide MTDP as a developer mode to encourage users choosing their preferred teacher model for distillation.

%
%


\section*{Funding}
City University of Hong Kong (Project 9229134) and Hong Kong Innovation and Technology Fund (ITF) [MRP/071/20X].

\section*{Funding}
City University of Hong Kong (Project 9678241, 9440274 and 7005453) and the Hong Kong Innovation and Technology Commission (InnoHK Project CIMDA).

\bibliographystyle{unsrt}  
\bibliography{references}

\end{document}